\documentclass[prl,aps,twocolumn,superscriptaddress]{revtex4}

\usepackage{epsfig}
\usepackage{graphicx,color}% Include figure files
\usepackage{dcolumn}% Align table columns on decimal point
\usepackage{bm}% bold math
\usepackage{amssymb}
\usepackage{ulem}

\begin{document}

%\preprint{Preprint}

\title{Mott transition in a metallic  liquid -- Gutzwiller molecular dynamics simulations}

\author{Gia-Wei Chern}

\affiliation{Theoretical Division, Los Alamos National Laboratory, Los Alamos, New Mexico 87545, USA}
\affiliation{Center for Nonlinear Studies, Los Alamos National Laboratory, Los Alamos, New Mexico 87545, USA}
\affiliation{Department of Physics, University of Virginia, Charlottesville, VA 22904, USA}

\author{Kipton Barros}

\affiliation{Theoretical Division, Los Alamos National Laboratory, Los Alamos, New Mexico 87545, USA}
\affiliation{Center for Nonlinear Studies, Los Alamos National Laboratory, Los Alamos, New Mexico 87545, USA}

\author{Cristian D. Batista}

\affiliation{Theoretical Division, Los Alamos National Laboratory, Los Alamos, New Mexico 87545, USA}
\affiliation{Center for Nonlinear Studies, Los Alamos National Laboratory, Los Alamos, New Mexico 87545, USA}

\author{Joel D. Kress}

\affiliation{Theoretical Division, Los Alamos National Laboratory, Los Alamos, New Mexico 87545, USA}

\author{Gabriel Kotliar} 

\affiliation{Department of Physics and Astronomy, Rutgers University, Piscataway, NJ 08854-8019, USA}

\date{\today}

\begin{abstract}
We present a formulation of quantum molecular dynamics that includes electron correlation effects via the Gutzwiller method. Our new scheme enables the study of the dynamical behavior of atoms and molecules with strong electron interactions. The Gutzwiller approach  goes beyond the conventional mean-field treatment of the  intra-atomic electron repulsion and captures crucial correlation effects such as band narrowing and electron localization. We use Gutzwiller quantum molecular dynamics to investigate the Mott transition in the liquid phase of a single-band metal and uncover intriguing structural and transport properties of the atoms. 
\end{abstract}

\maketitle

%The electron correlation effects also significantly affect static as well as dynamic properties of the ionic systems. Indeed, although the detailed mechanism is still under debate, it is commonly agreed that electron correlations play a crucial role in the volume collapse transition observed in some rare-earth metals at high pressures.

The  physics of Mott Hubbard correlations has been  intensely studied because of its conceptual relevance to many classes  of correlated materials. 
Previous studies have largely assumed fixed atom positions~\cite{Kadzielawa14}. To understand correlated electron physics within metallic liquids, it is imperative to include correlation effects also on the atomic dynamics. In this Letter we incorporate the Gutzwiller method into quantum molecular dynamics (QMD) to elucidate the basic questions of how proximity to the Mott transition affects correlation functions of the liquid and how it impacts  ionic and electronic transport.

The Gutzwiller variational wave function, together with the Gutzwiller approximation (GA)~\cite{gutzwiller,gutzwiller2,gutzwiller3} provide an efficient approach to correlated materials. The basic idea is  to apply an  operator to a Slater determinant, which reduces the probability amplitude of doubly occupied states. The optimum double occupancy probability is determined variationally. As in the mean-field approach, the GA retains the desirable feature of an effective single-particle picture. More importantly, it captures crucial correlation effects, such as bandwidth renormalization. For instance, the Brinkman-Rice theory of the Hubbard model~\cite{brinkman70}, which provided one of the first important steps towards our understanding of the Mott transition (MT), is based on the GA. Subsequently, the GA has been reformulated as the saddle point solution of a slave boson theory, leading to multiple generalizations and broadened applicability~\cite{kotliar86}.  Moreover, the GA can be combined with density functional theory (DFT)~\cite{hohenberg64,kohn65}. Indeed, the local density approximation combined with the GA has proven to be a powerful scheme for studying real  correlated metals~\cite{Lanata15,julien06,ho08,deng09,bunemann08,borghi14,julien15}.

It is well-known that strong intra-atomic Coulomb interaction ($U$) induces electronic localization, which can trigger iso-structural transitions with large volume collapse in f-electron {\it lattice systems}, such as metallic Ce and Pu  \cite{Lanata14,Lanata14b,Lanata15,Allen82,Lavagna82,Johansson74}.
Similarly, we expect that the corresponding Mott-Anderson transition in correlated {\it liquid metals} will induce drastic changes in  static and transport properties. 
%Although these electronic transport properties are well studied~\cite{Mott97,Belitz94}, much less is known about the atomic properties near a Mott-Anderson transition.
Our Gutzwiller QMD (GQMD) scheme elucidates the $U$-dependence of {\it both} the electronic and ionic transport coefficients. Specifically we develop a tight-binding (TB) QMD simulation coupled to a robust Gutzwiller solver. Tight-binding is much faster, although less accurate, than full DFT-based MD~\cite{TBMD}.
%Although our prototype implementation uses direct numerical diagonalization to solve the TB Hamiltonian, the proposed GA scheme %could instead be based upon linear-scaling methods. 
We demonstrate our approach by investigating the effect of intra-atomic Coulomb repulsion on the structural and dynamical properties of of the simplest  possible model Hamiltonian, which describes  a narrow $s$-band liquid metal. Besides the large volume expansion, we find that the strongly first order metal-insulator transition is accompanied by a drastic drop of the ionic self-diffusion coefficient.

We consider single-orbital atoms with an on-site Hubbard interaction $U$ in a tight-binding formulation:
\begin{eqnarray}
	\label{eq:H_total}
	\mathcal{H}_e &=&  \sum_{i\neq j} \sum_\sigma t(|\mathbf r_i - \mathbf r_j|) c^\dagger_{i, \sigma} c^{\;}_{j, \sigma} + U \sum_{i} n_{i, \uparrow} n_{i, \downarrow}  \nonumber \\
  	 & & +  \frac{1}{2} \sum_{i\neq j} \phi(|\mathbf r_i - \mathbf r_j|) + \sum_i \frac{ |\mathbf p_i|^2 }{2m}.
\end{eqnarray}
The first term is the electron hopping between neighboring atoms. The operator $c^\dagger_{i,\sigma}$ creates an electron  with spin $\sigma = \uparrow$, $\downarrow$ at the $i$-th atom. $n_{i,\sigma} = c^\dagger_{i, \sigma} c^{\;}_{i,\sigma}$ is the electron number operator and $\mathbf r_i$ is the position vector of $i$-th atom. $\phi(r)$ is the pairwise repulsive inter-atomic potential. The last term of (\ref{eq:H_total}) is the atomic kinetic energy ($m$ and $\mathbf p_i$ are the atomic mass and momentum, respectively). For simplicity, we assume that both the hopping and pair-potential scale exponentially with the interatomic distance: $t(r) = t_0 \exp(-r/\xi_1)$ and $\phi(r) = \phi_0 \exp(-r / \xi_2)$. In applications to real materials, these parameters are usually determined by fitting to bulk band-structure {\it ab initio} calculations or to experimental results~\cite{kwon94}. Our scheme does not depend on details of this parametrization. 

%The parameters $t_0$, $\phi_0$, and $\xi_{1,2}$ are chosen to mimic the energy curve of hydrogen atoms obtained in Ref.~\cite{kwon94}.

To efficiently include correlation effects induced by $U$, we adopt the GA and obtain the optimum many-electron wave function {\it at each time step} of the MD simulation. The  optimized wave function depends only on the instantaneous ionic configuration when we adopt  the Born-Oppenheimer approximation. Specifically, the ionic configuration $\{\mathbf r_i\}$ at each time step determines a tight-binding model parametrized by hopping amplitudes $t_{ij} = t(|\mathbf r_i - \mathbf r_j|)$. A Slater determinant  $|\Psi_0\rangle$ is obtained from the single-particle eigenstates of the TB Hamiltonian. The correlated many-electron wave function is approximated by $|\Psi_G \rangle = \prod_i \mathcal{P}_i |\Psi_0\rangle$, where $\mathcal{P}_i$ is the Gutzwiller operator. Within the GA, which is exact in the infinite dimension limit, the expectation value of the off-site term acquires a renormalization: $\langle \Psi_G | c^\dagger_{i, \sigma} c^{\;}_{j, \sigma} |\Psi_G \rangle = \mathcal{R}_{i, \sigma} \mathcal{R}_{j, \sigma} \langle \Psi_0 | c^\dagger_{i, \sigma} c^{\;}_{j, \sigma} | \Psi_0 \rangle$, where
\begin{eqnarray}
	\mathcal{R}_{i, \sigma} = \frac{ \sqrt{(1- \rho_{ii} + d_i)(\rho_{ii,\sigma} - d_i)} + \sqrt{d_i (\rho_{ii, \bar \sigma} - d_i)} }{\sqrt{ \rho_{ii,\sigma} (1-\rho_{ii,\sigma}) } }. \nonumber \\
\end{eqnarray}
Here $\rho_{ij, \sigma} = \langle \Psi_0 | c^\dagger_{i, \sigma} c^{\;}_{j, \sigma} | \Psi_0 \rangle$ is the single-particle density matrix, $\rho_{ii} = \rho_{ii, \uparrow} + \rho_{ii, \downarrow}$, and $d_i = \langle \Psi_G | n_{i, \uparrow} n_{i, \downarrow} | \Psi_G \rangle$ is the double occupancy probability at $i$-th atom. The  $\{d_i\}$ variables are treated as variational parameters to be determined by minimizing the Gutzwiller energy functional,
\begin{eqnarray}
	E_G = \frac{1}{2} \sum_{i\neq j}\sum_\sigma t_{ij} \, \mathcal{R}_{i,\sigma} \mathcal{R}_{j,\sigma}\, \rho_{ij,\sigma}+ U \sum_i d_i,
\end{eqnarray} 
Since we are interested in the high-temperature ($T$) liquid regime of  $\mathcal{H}_e$, we employ a finite-$T$ extension of the GA developed in Ref.~\cite{wang10}. The entropy correction due to the Gutzwiller operator is approximated by the lower bound  $\Delta S = \ln \langle \Psi_0 | \mathcal{P} | \Psi_0 \rangle$. Although this approach leads to an unphysical negative 
low-$T$ entropy, it nonetheless gives a good approximation in the high-$T$ regime of  interest. More importantly, the inclusion of this entropy correction for a half-filled Hubbard model on a lattice reproduces the
critical endpoint of the first order  metal-insulator transition line obtained with DMFT~\cite{wang10}. Within this finite-$T$ extension of the GA, the variational parameters $\{ d_i \}$ are obtained by minimizing the total free energy:
\begin{eqnarray}
	F_G &=& -k_B T \ln {\rm Tr} \, e^{-\beta \tilde \mathcal{H}_{\rm TB}} \\ 
	& & \!\! - \sum_i \left(e_i \ln\frac{e_i}{e_{i0}} + q_i \ln\frac{q_i}{q_{i0}} + d_i \ln \frac{d_i}{d_{i0}}\right). \nonumber
\end{eqnarray}
where $e_i$ and $q_i$ are the empty and single-occupancy probabilities, and the subscript $0$ denotes the  probabilities for the uncorrelated wave function. The first term is the free energy of non-interacting fermions, whose TB Hamiltonian is renormalized by~$\mathcal{R}_{i,\sigma}$. The second term arises from the correction $\Delta S$ of the Gutzwiller operators.
Per the Born-Oppenheimer approximation, we assume the electrons are always in thermal equilibrium.
Because strong correlations can produce a drastic suppression of the effective Fermi temperature,  a strongly  temperature dependent  free energy functional  is essential for  describing correlated electronic degrees of freedom with MD~\cite{Yin13}.

In modern implementations of the GA, both the Slater determinant $|\Psi_0 \rangle$ and the variational parameters $\{d_i \}$ are to be optimized through iterations~\cite{lanata12}. The optimization of $|\Psi_0\rangle$ corresponds to solving the single-particle density matrix $\rho_{ij,\sigma}$ of the renormalized TB Hamiltonian $\tilde \mathcal{H}_{\rm TB}$, while the parameters $\{ d_i \}$ are obtained by minimizing $F_G$. In our implementation of the GQMD,  these two optimizations are repeated until the iterations converge. An important criterion for convergence is to verify the constraint $\langle \Psi_G | n_{i, \sigma} | \Psi_G \rangle = \langle \Psi_0 | n_{i, \sigma} | \Psi_0 \rangle$. Once the optimal solution is obtained, we compute the forces acting on the ions,
\begin{eqnarray}
	\mathbf f_i = -\sum_{j,\sigma} \frac{\partial t(r_{ij})}{\partial \mathbf r_{i}} \mathcal{R}_{i,\sigma} \mathcal{R}_{j,\sigma}\,\rho_{ij,\sigma} - \sum_j \frac{\partial\phi(r_{ij})}{\partial \mathbf r_i}.
\end{eqnarray}
Given the forces, we integrate the ionic positions one timestep using the velocity Verlet method with Langevin noises~\cite{MD}.

%. \textcolor{red}{\sout{using the velocity Verlet method~\cite{MD}.} [We're actually using Langevin dynamics, not simple velocity Verlet, right?]}

\begin{figure}[b]
\includegraphics[width=0.99\columnwidth]{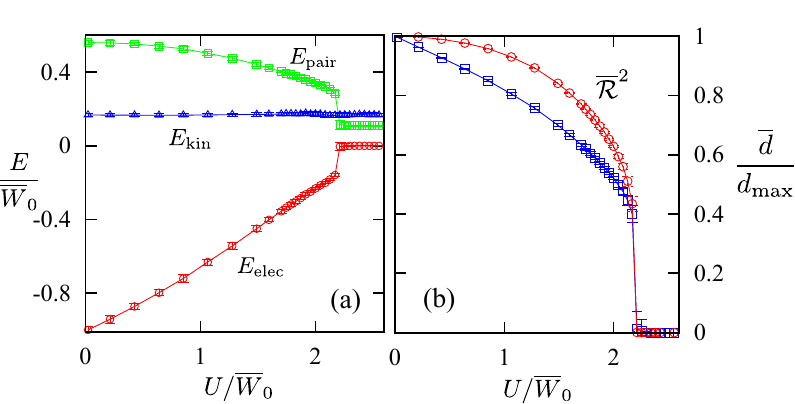}
\caption{(a) Average electronic energy $E_{\rm elec}$, pair-potential $E_{\rm pair}$, and kinetic energy $E_{\rm kin}$ as a function of $U$. The roughly constant kinetic energy is determined from the simulation temperature (not affected by $U$). (b) Average double occupancy ${\overline d}/d_{\rm max}$ and renormalization ${\bar \mathcal{R}}^2$ as a function of $U$. The maximum double-occupancy  is $d_{\rm max} = \langle n_{\uparrow} \rangle \langle n_{\downarrow} \rangle = 0.25$.
\label{fig:u_dep}}
\end{figure}

\begin{figure}
\includegraphics[width=0.99\columnwidth]{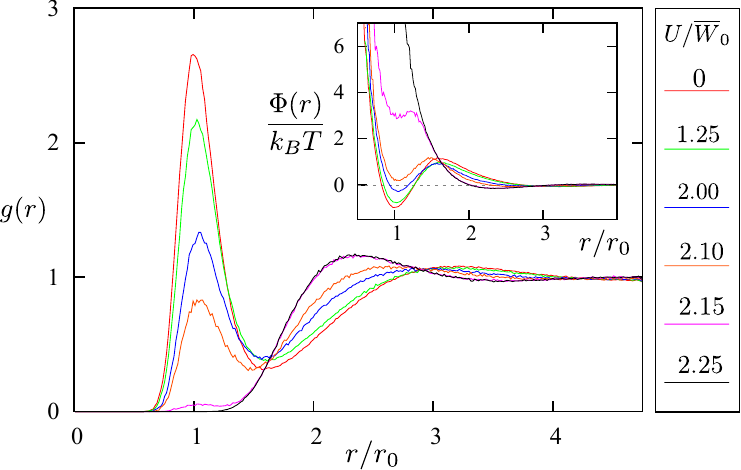}
\caption{Pair distribution function $g(r)$ obtained from GQMD for varying values of $U$. Here $d$ is the equilibrium distance between atoms in a molecule. The explicit value of $r_0$ is determined by the minimum of the energy curve $e(r) = -2 t(r) + \phi(r)$. The simulation temperature is $k_B T \sim 0.1 {\overline W}_0$. The inset shows the potential of mean force $\Phi(r) = -k_B T \ln g(r)$.
\label{fig:gr}}
\end{figure}

%{\em Mott metal-insulator transition}.
We now apply the GQMD to simulate the liquid phase of an $s$-band system, such as hydrogen at high temperatures. Since our main interest is the MT in the paramagnetic phase, we restrict ourselves to non-magnetic solutions. We use a constant volume $V$ and constant temperature MD simulation with $N = 100$ atoms. The temperature $T$ is kept constant by using a Langevin thermostat~\cite{MD} with a rather small damping $\gamma \sim 10^{-3} - 10^{-2}$ fs$^{-1}$. $V$ is determined from the average interatomic distance $r_s = (3V/4\pi N)^{1/3} \approx 1.6 r_0$, where $r_0$ is the equilibrium distance between two atoms in the molecular state (e.g. H$_2$). Here we will only consider the half-filled case with a  number of electrons $N_e = N$. The atoms are randomly distributed within the simulation box at the beginning of the simulation, and the system relaxes to equilibrium in 10000 to 50000 MD steps. The time step is 0.5 fs and the total trajectory simulations are of order $10^3$ fs. Our Gutzwiller solver is rather efficient; it takes an average of less than 10 iterations to reach convergence  in equilibrium, but the number of iterations can be as high as a few hundreds during the relaxation process.

Fig.~\ref{fig:u_dep}(a) shows the electron, pair-potential, and kinetic energy as a function of  $U$. All the energies are normalized to the average band energy $\overline W_0$ at $U=0$.
In equilibrium, the Langevin thermostat ensures that the kinetic energy per atom satisfies $E_{\rm kin} = \frac{3}{2} k_B T$.  Both the electronic energy $E_{\rm elec}$ and the pair-potential $E_{\rm pair}$ show a pronounced change at a critical value $U_c \approx  2.1\, \overline W_0$. The electronic energy here includes the binding energy of the TB Hamiltonian and the Hubbard interaction term. Above $U_c$, the electron energy vanishes identically indicating that the system enters a new phase with distinctly different electronic properties. The nature of this MT can be inferred from the $U$ dependence of the average renormalization and the double-occupancy probability shown in Fig.~\ref{fig:u_dep}(b). Here ${\overline\mathcal{R}}^2$ is used as an estimate of the electron bandwidth renormalization.  The averaged renormalization is close to one, ${\overline\mathcal{R}}^2 \approx 1$, for $U \ll \overline W_0$ and it quickly decreases to zero for $U \gtrsim U_c$, indicating a first-order transition. 
%The inset of Fig.~\ref{fig:u_dep}(b) shows the average double occupancy $\overline d$ as a function of $U/U_c$ for two different temperatures; note that $U_c$  is temperature dependent. 
The double-occupancy remains close to the uncorrelated value, $\overline d \approx \langle n_{\uparrow} \rangle \langle n_{\downarrow} \rangle \approx 0.25$, for small $U$, while it vanishes above $U_c$. 
%The higher $T$ transition seems to be smoother. 
The renormalization to zero of the effective bandwidth shows that the metal-insulator transition is driven by electronic localization, as also evidenced by the vanishing double occupation for $U > U_c$.  
%Our results provide the first QMD simulations of this Mott, or more precisely Brinkman-Rice, transition in a metallic phase of atomic fluids.

The radial pair distribution function $g(r)$ is the probability of finding an atom at a distance $r$ from a reference atom~\cite{MD}. At small $U$,   the $g(r)$ curves obtained from our GQMD simulations   exhibit a pronounced peak at the equilibrium distance $r_0$ of the binding energy $e(r) = -2 t(r) + \phi(r)$ (see Fig.~\ref{fig:gr}). This peak arises from the formation of quasi-dimer molecules in the liquid phase~\cite{collins95,mazzola14}. As $U$ increases, the dimer peak gradually disappears, while the second broader peak moves toward longer distances. The trend is consistent with the MT scenario in which increasing Coulomb repulsion suppresses the formation of a covalent bond as electrons become localized. The molecular peak disappears above  $U=U_c\simeq 2.1 {\bar W}_0$ and the distribution function only exhibits a broad peak at $r \sim 2.2\,r_0$. 

\begin{figure}
\includegraphics[width=0.99\columnwidth]{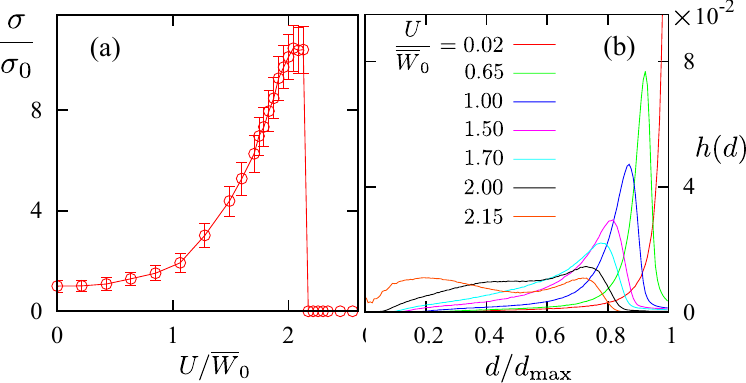}
\caption{(a) Dc conductivity $\sigma$ normalized to the value at $U=0$ as a function of $U$. (b) Distribution of the double-occupancy probability $d$ for  different $U$ values. The simulation temperature is $k_B T \sim 0.1 \overline W_0$.  
\label{fig:cond}}
\end{figure}

The metal-insulator transition is also demonstrated by the $U$-dependence of the dc conductivity $\sigma$ [Fig.~\ref{fig:cond}(a)] computed with the Kubo-Greenwood formula~\cite{ashcroft76,Desjarlais02}. The current operator has a simple form in the tight-binding basis~\cite{Mahan00}. While $\sigma$ vanishes above $U_c$, the sharp increase  when $U$ approaches $U_c$ from the metallic side is caused by the dimer dissociation, which is accelerated near the MT (see Fig.~\ref{fig:gr}). Similar to the case of doped semiconductors, each isolated atom (or monomer) introduces an electronic state inside the bonding-antibonding gap of the dimer spectrum~\cite{lenosky97,Ross96}.  According to Mott's approximation~\cite{Mott79}, $\sigma$ is proportional to the square of the density of states at the Fermi level, implying that it should also be proportional to the square of the monomer density $\rho_m$, as illustrated in a previous work~\cite{lenosky97}. The linear increase of $\rho_m$ with $U$ for $1 \lesssim U/{\bar {W_0}} \lesssim 2$ explains the quadratic increase of $\sigma(U)$ in the same interval. The combination of this effect with the rapid suppression of $\sigma$ at the MT leads to the rather sharp peak at 
$U_c$.

The histogram of double-occupancy probability $h(d)$ shown in Fig.~\ref{fig:cond}(b) exhibits an interesting bimodal distribution when $U$ approaches $U_c$. The two peaks of this bimodal distribution arise from the  dissociation of
dimers into monomers. The  sharper dimer peak  is always centered around higher $d/d_{\rm max}$ values 
because the hopping amplitude between the two atoms in a dimer is clearly larger than the average 
hopping amplitude between monomers. The monomer peak is much broader because of the broader distribution of
monomer-monomer and monomer-dimer distances relative to the distribution distances between two ions in the same dimer.

\begin{figure}
\includegraphics[width=0.95\columnwidth]{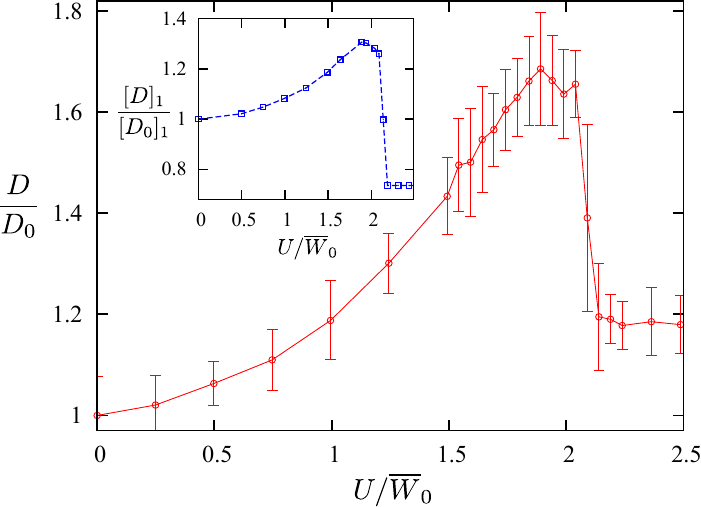}
\caption{Self-diffusion coefficient  $D$ as a function of $U$. $D_0$ is the diffusion constant for $U = 0$. The simulation temperature is $k_B T \sim 0.1 {\overline W}_0$. The inset shows the same self-diffusion coefficient, $[D]_1/[D_0]_1$, as approximated by the Chapman-Enskog theory for the effective potential $\Phi(r)$.
\label{fig:diffusion}}
\end{figure}

The coefficient of self-diffusion is an important measure of the dynamics of the liquid phase. It is computed from the velocity autocorrelation function~\cite{MD},
\[
D = \frac{1}{3 N} \sum_i \int_0^\infty \langle \mathbf v_i(t) \cdot \mathbf v_i(0) \rangle \, dt.
\]
Fig.~\ref{fig:diffusion} shows the normalized self-diffusion coefficient obtained from our GQMD simulations. The original increase of $D(U)$  is  related to the suppression of the molecular peak in $g(r)$.  At small $U$, the atoms form transient bound dimers, whose  larger effective mass  leads to smaller $D$ values. In parallel,  the simultaneous change of the effective two-atom
potential for increasing $U$ modifies the self-diffussion coefficient of the increasing number of monomers. 
To demonstrate that this  effect leads to
the drastic drop of $D$ at $U_c$, 
we compute the  Chapman-Enskog self-diffusion coefficient to first order~\cite{chapman70},
\begin{equation}
[D]_1= \frac{3}{8} \sqrt{\frac{ \pi k_B T}{m}} \left ( \frac{1}{n \Omega^{(1,1)}} \right ),
\end{equation} 
where  $n$ is the density and $\Omega^{(1,1)}$ is the collision integral for
diffusion  obtained from the effective two-atom potential of mean force $\Phi(r) = -k_B T \ln g(r)$ shown in the inset of Fig.~\ref{fig:gr}.
$\Phi(r)$ includes correlation effects self-consistently~\cite{kirkwood35,baalrud13}. 
In general, the effect of integrating out the  electrons cannot be reduced to a simple two-body effective interaction.
Nevertheless, because the coupling to the electronic degrees of freedom weakens near the MT (${\overline\mathcal{R}}^2$ is strongly suppressed),
we expect that the effective two-body potential provides a reasonable description in this ``weak-coupling" regime. 
Indeed, the result shown in the inset of  Fig.~\ref{fig:diffusion}  agrees quite well with the self-diffusion coefficient directly obtained from the GQMD simulation near the MT.
The same level of agreement is not obtained for small $U$ values because the formation of molecular states is not accounted for in this simplified analysis.

This calculation shows that the non-monotonic dependence of $D(U)$ arises from the change of $\Phi(r)$ with $U$.
$\Phi(r)$ is strongly attractive around $r \sim r_0$ for small $U$ values due to a large electronic contribution (see inset of Fig.~\ref{fig:gr}).
The corresponding potential barrier at $r \simeq 1.5 r_0$  increases the transport cross section in the energy range $E \lesssim k_B T$ most relevant for the collision integral. 
The height of this barrier decreases with $U$ because the attractive electronic component becomes weaker. 
The resulting suppression  of the transport cross section explains the corresponding increase of $D(U)$ shown in inset of  Fig.~\ref{fig:diffusion}.
This behavior changes drastically for $U \gtrsim U_c$ because of the discontinuous suppression of the electronic binding energy.
In this regime, the atoms interact only via the potential $\phi(r)$. 
The resulting  transport cross section is similar to the $U=0$ case for $E \lesssim k_B T$ and it is 
even higher for $E > k_B T$ (see inset of Fig.~\ref{fig:gr}). This  explains the drastic drop of $D$ at the MT.

%{\em Conclusion and Discussion}. 
%We have developed a novel GQMD method for dynamical simulations of  correlated materials. The forces acting on the atoms are computed from an optimized many-electron wave function that is determined %variationally within the GA. By applying method to a single-band system in the liquid phase, we demonstrate that strong electron correlations could significantly affect structural as 
%well as dynamical ionic properties, as in the case of solid state systems. 

MD simulations are widely used to understand fundamental properties of materials such as molecular structures, transport, phase transitions, and chemical reactions. Full quantum mechanical treatment of electron wavefunctions has the potential to greatly increase the predictive power of MD~\cite{car85}. For many functional materials, including transition metal and rare-earth compounds,  electron correlation effects are known to be crucial, yet are neglected even in  state-of-the-art MD simulations. 
%Conventional QMD studies use self-consistent mean-field approximations that fail to account for correlation induced phenomena, such as bandwidth renormalization and electron localization %transitions \cite{Lanata14,Lanata14b,Lanata15}. Our understanding of correlation effects in lattice systems has improved significantly thanks to sophisticated methods, such as dynamical %mean-field theory (DMFT), but much less is known about the effect of electron correlations on the atomic dynamics. 
%To shed light on this matter, it is imperative to efficiently incorporate electron correlation into MD simulations. 
Our implementation of the single-band case provides a proof of principle for including electron correlations in MD simulations. Although there are more accurate methods, e.g. MD  combined with variational quantum Monte Carlo~\cite{delaney06,attaccalite08} and path-integral QMD~\cite{margo96,pierleoni94}, for the single-band atoms such as hydrogen, the Gutzwiller MD scheme is the only method that can be feasibly generalized to multi-orbital correlated materials of primary interest, such as $d$ and $f$-electron systems,
as demonstrated recently by calculations of the equation of state of Pr and Pu~\cite{Lanata14c}.
%However, the number of variational parameters grows exponentially with the number of orbitals (it is of the order $10^3$ for $d$ electrons). An important step towards the multi-band %GQMD is to find a subset of parameters that capture the essential correlation effects of interest. 

Recent developments can increase the efficiency of GQMD. A dominant computational cost of electronic structure solvers is calculating the density matrix from the TB Hamiltonian. {Direct diagonalization has computational cost that scales cubically with the system size. The kernel polynomial method (KPM) can provide stochastic estimates of the electronic free energy~\cite{weisse06}. The gradient transformation of the KPM energy estimation procedure yields density matrix elements, as required by Gutzwiller, with computational cost that scales linearly with system size for both insulating and metallic systems~\cite{barros13,barros14}. 
Another future direction is generalizing the extended Lagrangian (XL) formalism~\cite{niklasson08,cawkwell12} to the GQMD self-consistency equations. 
%XL is a variation of Car-Parinello dynamics in which the %electron density oscillates about its self-consistent solution. With the XL method, the time-scale of the auxiliary dynamics can approach that of the physical dynamics while %retaining near perfect time-reversibility and typically requiring only a single density matrix build per time step~\cite{niklasson08,cawkwell12}. The Gutzwiller self-consistency %equations have similar form to the Kohn-Sham ones, suggesting that the Gutzwiller variational parameters may naturally be treated as XL dynamical variables. 
The integration of these technique into the GQMD seems very promising. 

\newpage

We acknowledge useful discussions with A.~Niklasson, Y. Motome, J.-P. Julien, and J.-X. Zhu. Work at LANL was carried out under the auspices of the U.S. DOE contract No. DE-AC52-06NA25396 through the LDRD program. 
G.~K. was supported by  DOE BES   DE-FG02-99ER45761.
The numerical simulations were performed using the CCS-7 Darwin cluster at LANL.

\end{document}